# Spin-valley locking, bulk quantum Hall effect and chiral surface state in a noncentrosymmetric Dirac semimetal BaMnSb$_2$

J.Y. Liu[1†], J. Yu[2†], J.L. Ning[1†], H.M. Yi[2+], L. Miao[3], L.J. Min[2,3], Y.F. Zhao[2], W. Ning[2], K.A. Lopez[3], Y.L. Zhu[2], T. Pillsbury[2], Y. B. Zhang[1], Y. Wang[2], J. Hu[4], H.B. Cao[5], F. Balakirev[6], F. Weickert[6], M. Jaime[6], Y. Lai[6], Kun Yang[7], J.W. Sun[1], N. Alem[3], V. Gopalan[3], C.Z. Chang[2], N. Samarth[2], C.X. Liu[2]*, R.D. Mcdonald[6]*, Z.Q. Mao[2,1]*

[1]Department of Physics and Engineering Physics, Tulane University, New Orleans, Louisiana 70118, USA

[2]Department of Physics, The Pennsylvania State University, University Park, Pennsylvania 16802, USA

[3]Department of Materials Science and Engineering, The Pennsylvania State University, University Park, Pennsylvania 16802, USA

[4]Department of Physics, University of Arkansas, Fayetteville, AR 72701, USA

[5]Oak Ridge National Laboratory, Oak Ridge, Tennessee 37831, USA

[6]Los Alamos National Laboratory, Los Alamos, New Mexico 87545, USA.

[7]Physics Department and National High Magnetic Field Laboratory, Florida State University, Tallahassee, FL 32306, USA

Spin-valley locking in the band structure of monolayers of MoS$_2$ and other group-VI dichalcogenides has attracted enormous interest, since it offers potential for valleytronic and optoelectronic applications. Such an exotic electronic state has sparsely been seen in bulk materials. Here, we report spin-valley locking in a bulk Dirac semimetal BaMnSb$_2$. We find valley and spin are inherently coupled for both valence and conduction bands in this material. This is revealed by comprehensive studies using first principle calculations, tight-binding and effective model analyses, angle-resolved photoemission spectroscopy and quantum transport measurements. Moreover, this material also exhibits a stacked quantum Hall effect. The spin-valley degeneracy extracted from the plateau height of quantized Hall resistivity is close to 2. This result, together with the observed Landau level spin splitting, further confirms the spin-valley locking picture. In the extreme quantum **limit, we have also observed a two-dimensional chiral metal at the side surface, which represents a novel topological quantum liquid. These findings establish BaMnSb$_2$ as a rare platform for exploring coupled spin and valley physics in bulk single crystals and accessing 3D interacting topological states.**

*Correspondence to:: zim1@psu.edu; cxl56@psu.edu; rmcd@lanl.gov

† These authors contributed equally.

I. INTRODUCTION

The combination of inversion symmetry breaking and spin-orbital coupling (SOC) in solid materials provides a route to achieve electronic states with spin polarization in the absence of magnetism. When this occurs in a material possessing valleys in its conduction and valence bands, spin polarization becomes valley dependent, thus creating a unique electronic state characterized by spin-valley locking. Such an electronic state was first realized in monolayers of group-VI transition metal dichalcogenides (TMDCs) such as $MoS_2$ [1–7]. Although the honeycomb lattice structure of bulk $2H$-$MoS_2$ is centrosymmetric, it becomes non-centrosymmetric on monolayer. In combination with inversion symmetry breaking, the SOC induced by the heavy element Mo leads to the spin splitting and spin-valley locking of the valence band [1–7]. The spin-valley locked electronic band structure of group-VI TMDC monolayers gives rise to topological valley transport properties such as photo-induced charge Hall effect, valley Hall effect, and spin Hall effect under zero magnetic field [1,8,9], as well as valley-dependent optical selection rule [1–4]. These exotic properties hold a great promise for potential applications in valleytronics, spintronics and optoelectronics [10].

Although inversion symmetry breaking and SOC can be found in many materials, it is challenging to identify candidate materials which can enable the combination of a valley degree of freedom with inversion symmetry breaking and SOC. Although spin-valley locking has been demonstrated on monolayers of group-VI TMDCs, it is very rarely seen in bulk materials. To the best of our knowledge, there are only two reported examples, i.e. $3R$-$MoS_2$ [11] and $2H$-$NbSe_2$ [12]. No materials beyond TMDCs have been reported to show spin-valley locking to date. In this article, we show a previously reported, three-dimensional Dirac semimetal $BaMnSb_2$ [13] features unique spin-valley locking. Contrasted with $MoS_2$ monolayer whose spin splitting is large in the valence band (0.15-0.46 eV) [1,11,14,15], but relatively small in the conduction band (1-50 meV) [16], $BaMnSb_2$ shows spin splitting of ~0.35eV in both conduction and valence bands, much larger than the Dirac gap (~50meV). In addition, we have also observed a three-dimensional (3D) quantum Hall effect (QHE) in this material, in contrast to the usual two-dimensional (2D) QHE. From the QHE, we have not only demonstrated the spin-valley degeneracy of 2, but also found evidence of a novel quantum liquid - 2D chiral surface metal present in the quantum Hall state within the quantum limit.

## II. METHODS

Single crystals of $BaMnSb_2$, $Ba(Mn_{0.9}Zn_{0.1})Sb_2$, and $(Ba_{0.9}Eu_{0.1})MnSb_2$ used in this study were grown using self-flux method [13]. Prior studies have shown pristine $BaMnSb_2$ is hole-doped [13]; our attempts of doping Zn to Mn sites and Eu to Ba sites aim to tune the chemical potential to reduce hole doping. The Zn-doped sample can indeed reach this goal, but the Eu doped sample turns out to be even more heavily hole doped. This is possibly because that the Eu-electron doping is less than the hole doping caused by Mn and Ba vacancies. Supplementary Table. S1 shows the comparison of carrier density, mobility and other parameters among these three types of samples.

Previous work reported $BaMnSb_2$ possesses a centrosymmetric tetragonal structure with the space group of I4/*mmm* [13,17]. As will be shown below, this material indeed involves weak orthorhombic distortion, which is hard to be resolved. We performed comprehensive structural analyses using scanning transmission electron microscopy (STEM), neutron scattering and optical second harmonic generation (SHG). The STEM sample was prepared using a focused ion beam (FIB) system. Two cross-sectional lamellas perpendicular to each other (oriented in [100] and [010] direction respectively) were lifted out from the same crystal. The atomic resolution high angle annular dark field (HAADF)-STEM images are taken with the Thermo Fisher Titan3 S/TEM equipped with a spherical aberration corrector. The neutron scattering experiment was performed at the Oak Ridge National Lab. The SHG polarimetry and imaging measurements were performed on a modified Witec Alpha 300S confocal Raman microscope in a far-field reflection geometry, using an 800-nm fundamental laser beam generated by a Spectra-Physics SOLSTICE ACE Ti: sapphire femtosecond laser system (pulse width ~ 100fs, repetition rate of 80 MHz). A λ/2 wave plate was utilized to control the polarization direction ($\theta$) of the incident field ($E^{\omega}$). The second harmonic field ($E^{2\omega}$) generated through the nonlinear optical process inside the sample was first spectrally filtered, and then decomposed either parallel or perpendicular to the polar axis by an analyzer and finally detected by a photo-multiplier tube (PMT). The schematic of our set up is shown in Fig. S1 of the supplementary material (SM).

The angle-resolved photoemission spectroscopy (ARPES) measurements were performed at Beamline 5-4 of the Stanford Synchrotron Radiation Light source using a Scienta DA30L electron analyzer. The energy and angle resolutions are ~ 9 meV and ~0.2°, respectively. The light spot size was set as $36um \times 26um$. The samples were cleaved and measured at 15 K in the high vacuum chamber (~ $5\times10^{-11}$ Torr). Magnetotransport properties of $BaMnSb_2$, $Ba(Mn_{0.9}Zn_{0.1})Sb_2$, and $(Ba_{0.9}Eu_{0.1})MnSb_2$ were measured using the National High Magnetic Field Pulsed Field Facility at Los Alamos National Laboratory and the physical property measurement system (Quantum Design). The Hall bar samples of $BaMnSb_2$ and $(Ba_{0.9}Eu_{0.1})MnSb_2$ were prepared using focused ion beam (FIB). To achieve homogeneous transport, before we cut the samples using FIB, we first deposited a layer of Au on both edges of the sample and then attached the leads to the Au layers using epoxy.

The density functional theory [18] calculations are carried out using the Vienna Ab-initio Simulation Package (VASP) [19]. The recently developed strongly-constrained and appropriately-normed (SCAN) meta-GGA [20,21] is used for its superior performance in description of different chemical bonds and transition metal compounds [20–25]. The projector-augmented wave (PAW) method [26,27] is employed to treat the core ion-electron interaction and the valence configurations are taken as Ba: $5s^26s^25p^6$, Mn: $3p^63d^64s^1$, and Sb: $5s^25p^3$, and an energy cutoff of 520 eV is used to truncate the plane wave basis. We use a Γ-centered 8×8×1 mesh K-space sampling for electronic self-consistent calculations within $10^{-6}$ eV per unit cell. Geometries of both the tetragonal phase and the zigzag distorted orthogonal phase of BaMnSb$_2$ were allowed to relax until the maximum ionic forces were below a threshold of 0.001 eV Å$^{-1}$.

## III. RESULTs

### A. Structure determination

As noted above, previous studies [13,17] suggest BaMnSb$_2$ has a layered tetragonal structure (*I*4/*mmm*), which is composed of alternative stacking of Sb square net layers and MnSb$_4$ tetrahedral layers, with the Ba layers sandwiched in between the Sb and MnSb$_4$ layers. The Sb layers generate Dirac fermions. Our structural analyses using STEM for BaMnSb$_2$ reveal a weak orthorhombic distortion, with the Sb layers forming zig-zag chains, as shown in Figs. 1a-1b. Figure. 1c shows simulated and experimental HAADF-STEM images along the [100]- and [010]-zone axes. These images indicate the Sb columns within the 2D Sb zig-zag chain layers are evenly spaced along the [010] axis (Fig. 1c upper), whereas the Sb columns' positions shift and form a dimmer-like profile along the [100] axis (Fig. 1c lower). The distortion is highlighted in the magnified experimental HAADF-STEM in Fig. 1d, which is inconsistent with a tetragonal structure where Sb columns should be evenly spaced along both the [100] and [010]- axes, but agrees well with an orthorhombic distortion where Sb atoms on the 2D Sb planes form zig-zag chains, as shown in Fig. 1b. The structure simulation based on first principle DFT calculations finds a non-centrosymmetric orthorhombic structure with the space group of *I2mm* can well describe the STEM images shown in Fig. 1c. Based on the structural parameters determined by the DFT calculations (supplementary Tab. S2), the STEM simulation [28,29] was performed and the simulated STEM images, shown in the left panels of Fig. 1c, match well with the experimental images in the right panels of Fig. 1c. Moreover, the calculated Sb2 shift relative to its position in the tetragonal structure is ~ 30pm (Fig. 1b), very close to the Sb2 shift (35pm) measured in the STEM image (Fig. 1d). The orthorhombic distortion is also evidenced by neutron scattering measurements (see Fig. S2 and section I.1 in SM).

Further, we also conducted optical second harmonic generation (SHG) polarimetry and microscopy and observed strong SHG signal (Figs. 1e and 1f), clearly demonstrating the inversion symmetry breaking in BaMnSb$_2$. By rotating the polarization of both the fundamental and the SHG light by 90°, we observed a switch of the contrast in the SHG imaging, indicating the existence of 90° domain walls in the sample. As shown in Fig. 1f, the SHG polarimetry taken in domains A and B can be modeled with *2mm* point group (2-fold axis is along the a-axis in crystal). Details of this model described in Section I.2 in SM reveal not only the twin walls visible on the surface but also suggest underlying orthorhombic domains below the surface.

### B. Band structure

With the *I2mm* orthorhombic structure, we have calculated the electronic band structure of BaMnSb$_2$. We find the four Dirac nodes along the $\Gamma - M$ line expected for the *I4/mmm* tetragonal structure phase are completely gaped out for the *I2mm* structure, while two gapped Dirac cones near the Fermi energy emerge at two different momenta, located symmetrically around the X-point along the X-M line [labeled by valley index $K_\pm$ in the inset of Fig. 2a]. The Dirac band dispersion is encircled with the red dashed box in Fig. 2a. The weak interlayer tunneling leads to the two-dimensional nature of the Dirac cones (see Sections I. 3 and 4 in SM). The strong SOC leads to large spin-splitting in each valley, as shown in Fig. 2b where spin projection (i.e. the $\langle S_z \rangle$ value) is color coded. Here it is worth noting that both conduction and valence bands show similar strength of spin splitting (~ 0.35 eV). As a result, the valley degrees of freedom is coupled to spin for both conduction and valence bands, in stark contrast with $MoS_2$ monolayer for which the spin splitting is large in the valence band [1,11,14,15], but small in the conduction band as noted above [16].

To verify the calculated band structure, we carried out ARPES measurements on $Ba(Mn_{0.9}Zn_{0.1})Sb_2$ crystals. The reason why we chose this sample for measurements is that it is the least hole doped. As shown in Fig. 2f, we find linearly dispersed bands cross the Fermi level at two momentum points on the $\overline{M}$-$\overline{X}$ line and these two crossing points are close to the $\overline{X}$ point. This can also be seen clearly in the constant energy map acquired through the intensity integration over the energy range from -15 meV to -5 meV, as shown in Fig. 2d where two point-like hole pockets neat $\overline{X}$ are readily discernable. The single crossing point along cut 1 in Fig. 2d and 2e is consistent with the picture that two Dirac cones at $K_\pm$ intersect near $\overline{X}$. In accordance with the calculated band structure (Fig. 2a), linear band crossing points should not appear near $\overline{Y}$. Our observation of crossing points near both $\overline{X}$ and $\overline{Y}$ is due to the existence of 90° domain walls in our samples as mentioned above. Fig. 2g plots the calculated band (dotted lines) along $\overline{MX}$ together with the measured band, from which we see a good agreement between theory and experiment. Although valley dependent spin polarization cannot be resolved directly from the current ARPES measurements, the total degeneracy of ~2 extracted from the QHE, which will be discussed below, demonstrates that spin degeneracy in each valley has been lifted, thus leading to the locking of spin and valley degrees of freedom.

To illustrate the physical origin of gapped Dirac cones in BaMnSb$_2$, we schematically show the band evolution with the orthorhombic distortion and SOC in Fig. 2c. In each Sb layer, there are two Sb atoms within one unit cell, labeled as Sb1 and Sb2 in Fig. 1b. The first principle calculations show that the conduction and valence bands around X (Y) mainly originate from the $p_x, p_y$ orbitals of the Sb2 (Sb1) atom. Therefore, we construct an 8-band tight-binding (TB) model with 2 sublattices, 2 orbitals ($p_x, p_y$) and 2 spin components, from which we can calculate energy dispersion that captures all qualitative features found in the first principles calculations (see Figs. S15, S16 and S17 in SM). To illustrate the band evolution from atomic orbitals with the TB model, we start from the *I*4/*mmm* structure without SOC, where the conduction and valence bands are close to each other around $\bar{X}$ and $\bar{Y}$ points, which are related by four-fold rotation along z. Band crossings between these two bands are found along $\bar{X}$-$\bar{M}$ ($\bar{Y}$-$\bar{M}$), protected by the mirror symmetry $m_x$ ($m_y$) that only flips $x$ ($y$) direction (Fig. S15). The zig-zag distortion that reduces *I*4/*mmm* to *I*2*mm* directly gaps out the band crossings near $K_\pm$ by breaking the mirror symmetry $m_x$ (Fig. S16), and the resultant two bands at $K_\pm$ are from the linear superposition of $p_x$ and $p_y$ orbitals, labeled as $p_\pm$ bands in Fig. 2c. On the other hand, a large gap opens at $\bar{Y}$ for a large distortion (Fig. S16 in SM). Finally, we add the on-site SOC, which removes the spin degeneracy of both conduction and valence bands on one valley (Fig. S17). Due to strong spin splitting, the bands $|p_+, \uparrow\rangle$ and $|p_-, \uparrow\rangle$ at $K_+$ ($|p_+, \downarrow\rangle$ and $|p_-, \downarrow\rangle$ at $K_-$) are pushed closer to each other and form the gapped Dirac cones with spin-valley locking as mentioned above and elaborated in SM (Fig. S18a).

With the understanding of the origin of Dirac cones, we further construct the effective Hamiltonian for the Dirac cone to compare with experiments (See Sections III. in SM). The effective model is constructed around $K_\pm$ for small $q = k - K_\pm$ based on the symmetry and TB model, which reads

$$h_\pm(q) = (E_0 \pm v_0 q_y)\tau_0\sigma_0 \pm v_1 q_x \tau_x \sigma_0 \pm v_2 q_y \tau_z \sigma_0 + [\pm E_1 \pm b_0(v_1^2 q_x^2 + v_2^2 q_y^2) + v_3 q_y]\tau_y \sigma_0 + \lambda_0 \tau_y \sigma_z . \quad (1)$$

Here $E_0$ and $v_{0,1,2}$ are material related parameters, $\lambda_0$ denotes on-site SOC, $E_1$ and $v_3$ are given by the distortion, and $\tau$ and $\sigma$ are Pauli matrices for orbital and spin indices, respectively. Without the $q$-quadratic term, the spin-independent part of the Hamiltonian is in the most general symmetry-allowed form to the first order of $q$, while the simple symmetry-allowed $q$-quadratic term is also included to explain the Landau Level splitting as discussed in the next section. From the energy dispersion $E_\alpha = (E_0 + \alpha v_0 q_y) \pm \sqrt{(v_1 q_x)^2 + (v_2 q_y)^2 + (\alpha E_1 + \alpha b_0(v_1^2 q_x^2 + v_2^2 q_y^2) + v_3 q_y \pm \lambda_0)^2}$ around $K_\alpha$ with $\alpha = \pm$, the energy gap at $q = 0$ reads $||E_1| - |\lambda_0||$, i.e. the difference between SOC ($|\lambda_0|$) and distortion strength ($|E_1|$). By choosing appropriate parameters, we find a good fitting for the energy dispersion between our effective model and the first principles calculation within the experimentally relevant Fermi energy ranges (see Fig. S18b). The effective Hamiltonian Eq. (1) explicitly demonstrates the existence of gapped Dirac cones in our system.

### C. Observation of bulk QHE and Landau level splitting

The spin-valley locking picture discussed above is further corroborated in our quantum transport measurements. We observed a bulk QHE in BaMnSb$_2$ single crystals due to its layered structure. The inset in Fig. 3e shows an optical image of a 6-electrode Hall bar sample made on a BaMnSb$_2$ single crystal with the thickness of 91 μm (denoted as B#1 below). Fig. 3a presents the longitudinal ($\rho_{xx}$) and Hall ($\rho_{xy}$) resistivity data measured on this sample at 1.4K. Both $\rho_{xx}$ and $\rho_{xy}$ exhibit Shubnikov–de Haas (SdH) oscillations starting from ~3T and the oscillation frequency $B_F$ obtained from their fast Fourier transform (FFT) analyses is ~18.9 T. When the magnetic field is above 5T, $\rho_{xy}$ displays clear plateau features while $\rho_{xx}$ reaches minima, implying the presence of bulk QHE in BaMnSb$_2$. The robust evidence for this QHE is given in Fig. 3b which plots $1/\rho_{xy}$ scaled by the step size of the successive $1/\rho_{xy}$ plateaus (i.e. $1/\rho_{xy}^0$, see Fig. S3) as a function of $B_F/B$. When $\rho_{xx}$ reaches minima, $\rho_{xy}^0/\rho_{xy}$ is clearly quantized to half-integer numbers $\gamma = j+1/2$ ($j$, non-negative integer number), which corresponds to the half integer normalized filling factor given by $B_F/B$ at the $\rho_{xx}$ minima. All these signatures can be attributed to a stacked QHE, with each 2D Sb layer sandwiched by the Ba-MnSb4-Ba insulating slabs (Fig. 1a) acting as a quantum Hall layer. The non-trivial Berry phase (~0.97π) determined from the Landau level (LL) index fan diagram in Fig. 3c confirms the relativistic characteristic of quasi-particles, consistent with the previous reports [13,30]. The QHE observed in sample B#1 persists up to $T = 50$K, as shown by the temperature dependences of $\rho_{xx}$ and $\rho_{xy}$ in Figs. 3e and 3f.

Although this bulk QHE bears some similarity with the previously-reported bulk QHE in EuMnBi$_2$ [31] where the 2D Bi square-net layers act as QHE layers, the bulk QHE in BaMnSb$_2$ displays distinct features which are absent in EuMnBi$_2$. The QHE in EuMnBi$_2$ is driven by the canted antiferromagnetic (AFM) order produced by the Eu sub-lattice, which reduces interlayer coupling significantly. However, such a canted AFM state exists only in a limited field range (5-22 T), which renders the primary quantum Hall state within the quantum limit inaccessible. In contrast, for BaMnSb$_2$, there is no such a canted AFM order and the presence of its bulk QHE is attributed to intrinsically weak interlayer coupling. With this advantage, we observe the $\rho_{xy}$ plateau within the quantum limit near 35T where only the zero$^{th}$ Landau level is filled (i.e. the $\rho_{xy}$ plateau with $\gamma=B_F/B =1/2$ shown in Fig. 3b). Although the $\rho_{xx}$ minimum does not reach zero at $B_F/B =1/2$, the Hall angle ($\rho_{xy}/\rho_{xx}$) reaches ~ 2500% (Fig. S5) and the longitudinal conductivity $\sigma_{xx}$ is close to zero (Fig. S6). The non-zero value of $\rho_{xx}$ at the quantum Hall states can be attributed to the combination of the imperfect QHE induced by defects and the bulk conduction caused by the trivial bands crossing the Fermi level near $\Gamma$ (Fig. 2a). The trivial bands near $\Gamma$ are clearly resolved in our ARPES measurements which were conducted using the photon energy of 25 eV (see Fig. S4 in SM).

In the $\gamma=1/2$ quantum Hall state, we have also observed clear signatures of Landau level (LL) splitting. For sample B#1, this is manifested by the peak splitting in d$^2\rho_{xx}$/dB$^2$ near 1/B =0.05 in Fig. 3d. Such a LL splitting is much more clearly resolved in the (Ba$_{0.9}$Eu$_{0.1}$)MnSb$_2$ sample (E#1) which has a much higher mobility (5040 cm$^2$V$^{-1}$s$^{-1}$) than sample B#1 (1645 cm$^2$V$^{-1}$s$^{-1}$) and slightly higher carrier density. As seen in Fig. 4a, Fig. S7 and Fig. S8a, although the carrier density of sample E#1 (~1.48×10$^{12}$ cm$^{-2}$) is higher than that of sample B#1 (9.1×10$^{11}$ cm$^{-2}$), we can still observe the quantized $\rho_{xy}$ plateau with $\gamma=1/2$ in the quantum limit above 40T (Fig. S8a). The $\rho_{xx}$ of this sample exhibits striking splitting (marked by the arrows in Fig. 4a) near $B_F/B =1$ where $\rho_{xy}$ displays a steep increase.

Since the bulk QHE in BaMnSb$_2$ is attributed to the parallel transport of 2D Sb layers stacked along the c-axis, the spin valley degeneracy per layer $s$ can be derived from the step size between the successive 1/$\rho_{xy}$ plateaus (i.e. 1/$\rho_{xy}^0$ ) via the equation 1/$\rho_{xy}^0 = sZ^*(e^2/h)$ where $Z^*$ represents the number of layers per unit length [31]. Given each unit cell of BaMnSb$_2$ contains two Sb conducting layers (see Fig. 1a), $Z^* = 1/(c/2)$. Hence $1/\rho_{xy}^0 = s(2/c)(e^2/h)$. The values of $s$ for samples B#1 and E#1 derived from $1/\rho_{xy}^0$ is 1.5 and 2.3 respectively. Measurements on another Eu-doped sample (E#3) yields $s = 2.2$ (Fig. S9 and Fig. S8b). For Ba(Mn$_{0.9}$Zn$_{0.1}$)Sb$_2$(Z#1), since its carrier density is about one half of that of sample B#1 (see Tab. S1), its quantum Hall state of $\gamma = 3/2$ can be reached at about 7T (Fig. 4d and Fig. S8c) and the $s$ value estimated for this sample is 1.5. The $s$ values of sample E#1 ($s = 2.3$) and E#3 ($s = 2.2$) agree well with the expected value of $s = 2$ for the spin-valley locking electronic structure shown in Fig. 2b. The small deviation of the experimental values from 2 for these two samples should be due to the errors in the measurements of sample thickness ($|\Delta l/l| \leq 12\%$). The smaller $s$ values for samples B#1 ($s=1.5$) and Z#1 ($s=1.5$) can be ascribed to the inhomogeneous transport due to dead layers and/or imperfect contacts, which is not rare in stacked quantum Hall systems [31]. Additionally, it is worth mentioning that although twin domains are present in BaMnSb$_2$ due to its orthorhombic structure, it should not affect the value of $s$, since it is band structure that determines $s$. Domain walls could cause additional scattering to electron waves. Domain wall scattering in BaMnSb$_2$ should be weak, since strong SdH oscillations and QHE would not be observed if domain wall scattering was strong.

To compare with the QHE observed in experiments, we derive the LLs of Eq.(1) with Peierls substitution $\boldsymbol{q} \rightarrow \boldsymbol{q} + \frac{e}{\hbar}\boldsymbol{A}$ for the external magnetic field $\boldsymbol{B} = \boldsymbol{\nabla} \times \boldsymbol{A} = (0,0,B)$ (See Sec. IV in SM). Given $\lambda_0 E_1 < 0$ from the fitting, the analytical solutions to the leading order are given by $\epsilon_0^{+,\uparrow} = E_0 + E_1 + \lambda_0$ and $\epsilon_0^{-,\downarrow} = E_0 - E_1 - \lambda_0$ for the zero$^{\text{th}}$ LLs. Except for the zero$^{\text{th}}$ LLs, all the other LLs are doubly degenerate due to the valley degeneracy as schematically shown in the left panel of Fig. 4e, which accounts for the experimentally observed half-integer filling factor in quantum Hall states (Fig. 3b). As noted in Fig. 4a, the clear LL splitting has been observed in sample E#1 near 30T, which may originate from the Zeeman term $\mu_B B \sigma_z$ and/or the $\boldsymbol{q}$-quadratic terms (Fig. 4e). Since the strong SOC splits the spin degeneracy and locks spins to valleys, the $\boldsymbol{q}$-quadratic terms may also lift the spin-valley degeneracy and cause the LL splitting. Then, we estimate the magnitude of the Zeeman and $\boldsymbol{q}$-quadratic contributions to the LL splitting, and find that the energy scale of $\boldsymbol{q}$-quadratic terms is around 0.3 ($B$/Tesla) meV, much larger than that of Zeeman term $\mu_B B \sim 0.05$ ($B$/Tesla) meV (assuming the $g$-factor to be 2 here). Thus, this estimate suggests that the $\boldsymbol{q}$-quadratic terms play the major role in inducing the LL splitting. To experimentally evaluate the Zeeman term contribution, we have also measured the angular dependence of the SdH oscillations in interlayer resistance for an Eu-doped sample. As discussed in section I.5 in the SM, the results from these measurements (Fig. S13) indeed suggest a weak Zeeman effect and its $g$ factor being much less than that of EuMnBi$_2$ ($g$ = 9.8(4)) [32], which provides support for our theoretical assumption of $g$ = 2.

The LLs derived from the Hamiltonian with both $\boldsymbol{q}$-quadratic terms and Zeeman coupling are plotted in Fig. 4f and the corresponding density of states is shown in the bottom panel of Fig. 4f, from which we indeed can see the LL splitting for the magnetic field near 30T, as shown by the blue line in Fig. 4f, in good agreement with the experimental observation shown in Fig. 4a (Note that we adopted the carrier density of sample E#1 in our calculations in order to compare the LL splitting with this sample). With increasing the disorder broadening strength [33], the LL splitting disappears and only odd number filling of LLs can be found (see the red curve in Fig. 4f), consistent with the experimental observation of suppressed LL's splitting in a sample (#B1) with lower mobility (Fig. 3a). The consistency between theory and experiment in the LL splitting, together with the $s \sim 2$ value estimated from the $1/\rho_{xy}$ plateau height, provide strong evidence for the spin-valley locking in BaMnSb$_2$.

### D. 2D chiral surface metal in the quantum Hall state within the quantum limit

We have also performed measurements on the out-of-plane resistance ($R_{zz}$) as a function of magnetic field at various temperatures using another Eu-doped sample (E#2) with a similar carrier density as sample E#1. The $R_{zz}$ data of sample #E2 are presented in Fig. 4b. We find $R_{zz}$ exhibits a distinct plateau at 0.7K in the quantum Hall state within the quantum limit. Although the extent of the plateau is reduced in field range with increasing temperature, the extrapolated $R_{zz}$ values near 60T appears to saturate to a constant for $T < 20$ K. This trend is clearly manifested by the temperature dependence of $R_{zz}$ at 58.5 T (Fig. 4c) and is inconsistent with the generally-expected quantum Hall insulating state, but expected of the 2D chiral surface state in stacked QHE layers [34]. Previous studies on semiconductor superlattices have shown that the 2D chiral surface state dominates the $z$-axis transport while the bulk is at the quantum Hall insulating state and this leads the $z$-axis conductivity to be temperature independent below 0.2 K [35]. Given that bulk BaMnSb$_2$ is a system with stacked QHE layers as discussed above, it is reasonable to attribute the tendency of $R_{zz}$ saturation below 20K to the 2D chiral surface state. The chiral surface states surviving up to 20K offers an excellent opportunity to investigate its underlying physics. The 2D chiral surface state in the QHE represents a novel quantum liquid comprised of gapless excitations. Although 3D stacked QHE has been observed in several single crystal materials such as EuMnBi$_2$ [31] and ZrTe$_5$ [36], the surface chiral metal has not been reported for any of them. Moreover, it is also worth mentioning that the recently-reported bulk QHE in Cd$_3$As$_2$ [37] is not a stacked QH system, but arises from the Weyl orbital comprised of the surface Fermi arc on the opposite surfaces of the sample and the 1D chiral LLs in the bulk [38].

### E. DISCUSSIONS

The above discussions have shown the Dirac cone at $K_+$ carries only upward spins, while the other one at $K_-$ has downward spins (see Fig. 2b and the inset to Fig. 2a). Although this shares some similarity with the spin-valley locking of the monolayers group-VI TMDCs [1–6], the spin-valley locking in BaMnSb$_2$ exhibits several distinct features. First, the SOC-induced spin splitting at each valley (~0.35 eV) is much larger than the energy gap of Dirac cones (~50 meV) in BaMnSb$_2$, while the TMDC monolayer (e.g. MoS$_2$) is in the opposite limit, with the energy gap (~2eV) being much larger than the SOC-induced spin splitting (~0.15-0.46 eV for the valence band) [1,11,14–16]. The smaller band gap of Dirac cones compared to the SOC strength in our system implies that the Berry curvature for spin up state (spin down state) is more concentrated around the $K_+$ ($K_-$) valley, which can possibly lead to a more strongly coupled valley-spin Hall effect, as compared to TMDC monolayers and bulk 3R-MoS$_2$ [11]. Secondly, the smaller band gap around tens of meV in BaMnSb$_2$ suggests that the optical probe of valley-spin physics will appear in THz frequency regime, rather than the visible light regime. Additionally, thanks to the weak interlayer tunneling (see section I.4 in SM), the valley-spin locking remains in bulk single crystals, contrasted with TMDC materials where the presence of inversion symmetry in bulk material (2H phase) or the films with even number of layers will obliterate the valley-spin physics. Finally, the smaller Dirac gap also suggests that the system is close to a topological phase transition and recent theory has suggested that the piezoelectric coefficient will vary discontinuously across this transition [39]. Thus, BaMnSb$_2$ will provide a material platform to test this theoretical prediction. Therefore, BaMnSb$_2$ offers a rare opportunity to explore novel spin-valley locking physics, as well as topological phase transition, in bulk materials.

We also emphasize that bulk quantum Hall systems are of particular interest, as exemplified by the observation of novel fractional QHE states in bilayer graphene [40]. Theory has also predicted that if a 3D fractional QHE can be realized in the intermediate tunneling regime of a layered material where the interlayer tunneling strength is on the same order of Coulomb energy [41], it can support both new e/3 fermionic quasiparticles capable of freely propagating both along and between layers, as well as new gapless neutral collective excitation modes, i.e. emergent "photon" modes. BaMnSb$_2$ may serve as a playground to test these predictions.

## IV. CONCLUSIONS

Through structural analyses using STEM, SHG and first principle calculations, we find BaMnSb$_2$ possesses a non-centrosymmetric orthorhombic structure with the space group of *I2mm*. From the combined efforts of first principle band structure calculations, tight-binding and effective model analyses, ARPES and transport measurements, we have demonstrated that the interplay among inversion symmetry breaking, SOC and valley degree of freedom in BaMnSb$_2$ results in a unique electronic state with spin-valley locking. One distinct feature of the spin-valley locking in bulk BaMnSb$_2$ is that its SOC-induced spin splitting is much greater than the Dirac gap, which may leads to distinct topological valley transport properties, As such, BaMnSb$_2$ provides a rare opportunity to study coupled spin and valley physics in bulk single crystals. In addition, we also find strong evidence for the surface chiral metal state previously predicted for stacked quantum Hall systems.

**Note:** Our original manuscript was posted on arXiv in July 2019 (J.Y. Liu et al. arXiv:1907.06318). Over the course of revising our manuscript, we noted Sakai *et al*. also reported the study of QHE and spin-valley coupling of BaMnSb$_2$ in a manuscript posted on arXiv in Jan. 2020 (Sakai et al., arXiv:2001.08683) [42].

**ACKNOELEDGMENT**

We thank Jainendra Jain for insightful comments. This work was supported by the US Department of Energy under grants DE-SC0019068 and DE-SC0014208 (support for personnel, sample synthesis, high field measurements and data analyses); a part of sample synthesis and high-field measurements was supported by the U.S. Department of Energy under EPSCoR grant No. DE-SC0012432 with additional support from the Louisiana Board of Regents. Work at the National High Magnetic Field Laboratory was supported by National Science Foundation (NSF) DMR-1644779, the State of Florida, and the U.S. Department of Energy (DOE). F.B & M.J. acknowledge support from the DOE BES `Science of 100 T'program, RDM & YL acknowledges support from the Center for the Advancement of Topological Semimetals, an Energy Frontier Research Center funded by the U.S. DOE, Office of Basic Energy Sciences. R.D.M. also acknowledges support from the LANL LDRD DR20160085 'Topology and Strong Correlations' for the start of this work. The preliminary ARPES experiment was carried out at the 2DCC-MIP supported by NSF cooperative agreement DMR 1539916. KY's work is supported by National Science Foundation grants DMR- 1932796 and DMR-1644779. JH's Work is supported by the US Department of Energy, Office of Science, Basic Energy Sciences under grant DE-SC0019467. L.M., K.A.L., V.G. and N.A.'s work is supported by the Penn State Center for Nanoscale Science, an NSF MRSEC under the grant number DMR-1420620. C.X. L and J.Y acknowledge the support of the U.S. Department of Energy (Grant No.~DESC0019064) for the development of the theoretical model, and also the support from the Office of Naval Research (Grant No. N00014-18-1-2793) and Kaufman New Initiative research grant KA2018-98553 of the Pittsburgh Foundation. C. Z. C. acknowledges the support from the NSF-CAREER award (DMR-1847811) and the Gordon and Betty Moore Foundation's EPiQS Initiative (GBMF9063 to C. Z. C.).

**References:**

[1] D. Xiao, G.-B. Liu, W. Feng, X. Xu, and W. Yao, *Coupled Spin and Valley Physics in Monolayers of $MoS_2$ and Other Group-VI Dichalcogenides*, Phys. Rev. Lett. **108**, 196802 (2012).

[2] K. F. Mak, K. He, J. Shan, and T. F. Heinz, *Control of Valley Polarization in Monolayer $MoS_2$ by Optical Helicity*, Nature Nanotech **7**, 494 (2012).

[3] H. Zeng, J. Dai, W. Yao, D. Xiao, and X. Cui, *Valley Polarization in $MoS_2$ Monolayers by Optical Pumping*, Nature Nanotech **7**, 490 (2012).

[4] T. Cao, G. Wang, W. Han, H. Ye, C. Zhu, J. Shi, Q. Niu, P. Tan, E. Wang, B. Liu, and J. Feng, *Valley-Selective Circular Dichroism of Monolayer Molybdenum Disulphide*, Nat Commun **3**, 1 (2012).

[5] A. M. Jones, H. Yu, N. J. Ghimire, S. Wu, G. Aivazian, J. S. Ross, B. Zhao, J. Yan, D. G. Mandrus, D. Xiao, W. Yao, and X. Xu, *Optical Generation of Excitonic Valley Coherence in Monolayer $WSe_2$*, Nature Nanotechnology **8**, 634 (2013).

[6] X. Xu, W. Yao, D. Xiao, and T. F. Heinz, *Spin and Pseudospins in Layered Transition Metal Dichalcogenides*, Nature Physics **10**, 343 (2014).

[7] G. Aivazian, Z. Gong, A. M. Jones, R.-L. Chu, J. Yan, D. G. Mandrus, C. Zhang, D. Cobden, W. Yao, and X. Xu, *Magnetic Control of Valley Pseudospin in Monolayer $WSe_2$*, Nature Physics **11**, 148 (2015).

[8] K. F. Mak, K. L. McGill, J. Park, and P. L. McEuen, *The Valley Hall Effect in $MoS_2$ Transistors*, Science **344**, 1489 (2014).

[9] M. Onga, Y. Zhang, T. Ideue, and Y. Iwasa, *Exciton Hall Effect in Monolayer $MoS_2$*, Nature Materials **16**, 1193 (2017).

[10] J. R. Schaibley, H. Yu, G. Clark, P. Rivera, J. S. Ross, K. L. Seyler, W. Yao, and X. Xu, *Valleytronics in 2D Materials*, Nature Reviews Materials **1**, 1 (2016).

[11] R. Suzuki, M. Sakano, Y. J. Zhang, R. Akashi, D. Morikawa, A. Harasawa, K. Yaji, K. Kuroda, K. Miyamoto, T. Okuda, K. Ishizaka, R. Arita, and Y. Iwasa, *Valley-Dependent Spin Polarization in Bulk $MoS_2$ with Broken Inversion Symmetry*, Nature Nanotechnology **9**, 611 (2014).

[12] L. Bawden, S. P. Cooil, F. Mazzola, J. M. Riley, L. J. Collins-McIntyre, V. Sunko, K. W. B. Hunvik, M. Leandersson, C. M. Polley, T. Balasubramanian, T. K. Kim, M. Hoesch, J. W. Wells, G. Balakrishnan, M. S. Bahramy, and P. D. C. King, *Spin–Valley Locking in the Normal State of a Transition-Metal Dichalcogenide Superconductor*, Nature Communications **7**, 11711 (2016).

[13] J. Liu, J. Hu, H. Cao, Y. Zhu, A. Chuang, D. Graf, D. J. Adams, S. M. A. Radmanesh, L. Spinu, I. Chiorescu, and Z. Mao, *Nearly Massless Dirac Fermions Hosted by Sb Square Net in BaMnSb$_2$*, Scientific Reports **6**, 30525 (2016).

[14] Y. Zhang, T.-R. Chang, B. Zhou, Y.-T. Cui, H. Yan, Z. Liu, F. Schmitt, J. Lee, R. Moore, Y. Chen, H. Lin, H.-T. Jeng, S.-K. Mo, Z. Hussain, A. Bansil, and Z.-X. Shen, *Direct Observation of the Transition from Indirect to Direct Bandgap in Atomically Thin Epitaxial MoSe$_2$*, Nature Nanotechnology **9**, 111 (2014).

[15] J. A. Miwa, S. Ulstrup, S. G. Sørensen, M. Dendzik, A. G. Čabo, M. Bianchi, J. V. Lauritsen, and P. Hofmann, *Electronic Structure of Epitaxial Single-Layer MoS$_2$*, Phys. Rev. Lett. **114**, 046802 (2015).

[16] K. Marinov, A. Avsar, K. Watanabe, T. Taniguchi, and A. Kis, *Resolving the Spin Splitting in the Conduction Band of Monolayer MoS$_2$*, Nature Communications **8**, 1938 (2017).

[17] G. Cordier and H. Schäfer, *Darstellung Und Kristallstruktur von BaMnSb$_2$, SrMnBi$_2$ Und BaMnBi$_2$ / Preparation and Crystal Structure of BaMnSb$_2$, SrMnBi$_2$ and BaMnBi$_2$*, Zeitschrift Für Naturforschung B **32**, 383 (1977).

[18] W. Kohn and L. J. Sham, *Self-Consistent Equations Including Exchange and Correlation Effects*, Phys. Rev. **140**, A1133 (1965).

[19] G. Kresse and J. Furthmüller, *Efficient Iterative Schemes for Ab Initio Total-Energy Calculations Using a Plane-Wave Basis Set*, Phys. Rev. B **54**, 11169 (1996).

[20] J. Sun, R. C. Remsing, Y. Zhang, Z. Sun, A. Ruzsinszky, H. Peng, Z. Yang, A. Paul, U. Waghmare, X. Wu, M. L. Klein, and J. P. Perdew, *Accurate First-Principles Structures and Energies of Diversely Bonded Systems from an Efficient Density Functional*, Nature Chemistry **8**, 831 (2016).

[21] J. Sun, A. Ruzsinszky, and J. P. Perdew, *Strongly Constrained and Appropriately Normed Semilocal Density Functional*, Phys. Rev. Lett. **115**, 036402 (2015).

[22] J. Sun, B. Xiao, Y. Fang, R. Haunschild, P. Hao, A. Ruzsinszky, G. I. Csonka, G. E. Scuseria, and J. P. Perdew, *Density Functionals That Recognize Covalent, Metallic, and Weak Bonds*, Phys. Rev. Lett. **111**, 106401 (2013).

[23] D. A. Kitchaev, H. Peng, Y. Liu, J. Sun, J. P. Perdew, and G. Ceder, *Energetics of MnO$_2$ Polymorphs in Density Functional Theory*, Phys. Rev. B **93**, 045132 (2016).

[24] H. Peng and J. P. Perdew, *Synergy of van Der Waals and Self-Interaction Corrections in Transition Metal Monoxides*, Phys. Rev. B **96**, 100101 (2017).

[25] J. W. Furness, Y. Zhang, C. Lane, I. G. Buda, B. Barbiellini, R. S. Markiewicz, A. Bansil, and J. Sun, *An Accurate First-Principles Treatment of Doping-Dependent Electronic Structure of High-Temperature Cuprate Superconductors*, Communications Physics **1**, 1 (2018).

[26] P. E. Blöchl, *Projector Augmented-Wave Method*, Phys. Rev. B **50**, 17953 (1994).

[27] G. Kresse and D. Joubert, *From Ultrasoft Pseudopotentials to the Projector Augmented-Wave Method*, Phys. Rev. B **59**, 1758 (1999).

[28] C. Ophus, *A Fast Image Simulation Algorithm for Scanning Transmission Electron Microscopy*, Adv Struct Chem Imag **3**, 13 (2017).

[29] A. Pryor, C. Ophus, and J. Miao, *A Streaming Multi-GPU Implementation of Image Simulation Algorithms for Scanning Transmission Electron Microscopy*, Advanced Structural and Chemical Imaging **3**, 15 (2017).

[30] S. Huang, J. Kim, W. A. Shelton, E. W. Plummer, and R. Jin, *Nontrivial Berry Phase in Magnetic BaMnSb$_2$ Semimetal*, PNAS **114**, 6256 (2017).

[31] H. Masuda, H. Sakai, M. Tokunaga, Y. Yamasaki, A. Miyake, J. Shiogai, S. Nakamura, S. Awaji, A. Tsukazaki, H. Nakao, Y. Murakami, T. Arima, Y. Tokura, and S. Ishiwata, *Quantum Hall Effect in a Bulk Antiferromagnet EuMnBi2 with Magnetically Confined Two-Dimensional Dirac Fermions*, Science Advances **2**, e1501117 (2016).

[32] H. Masuda, H. Sakai, M. Tokunaga, M. Ochi, H. Takahashi, K. Akiba, A. Miyake, K. Kuroki, Y. Tokura, and S. Ishiwata, *Impact of Antiferromagnetic Order on Landau-Level Splitting of Quasi-Two-Dimensional Dirac Fermions in EuMnBi$_2$*, Phys. Rev. B **98**, 161108 (2018).

[33] E. G. Novik, A. Pfeuffer-Jeschke, T. Jungwirth, V. Latussek, C. R. Becker, G. Landwehr, H. Buhmann, and L. W. Molenkamp, *Band Structure of Semimagnetic Hg$_{1-y}$Mn$_y$Te Quantum Wells*, Phys. Rev. B **72**, 035321 (2005).

[34] L. Balents and M. P. A. Fisher, *Chiral Surface States in the Bulk Quantum Hall Effect*, Phys. Rev. Lett. **76**, 2782 (1996).

[35] D. P. Druist, P. J. Turley, K. D. Maranowski, E. G. Gwinn, and A. C. Gossard, *Observation of Chiral Surface States in the Integer Quantum Hall Effect*, Phys. Rev. Lett. **80**, 365 (1998).

[36] F. Tang, Y. Ren, P. Wang, R. Zhong, J. Schneeloch, S. A. Yang, K. Yang, P. A. Lee, G. Gu, Z. Qiao, and L. Zhang, *Three-Dimensional Quantum Hall Effect and Metal–Insulator Transition in ZrTe$_5$*, Nature **569**, 537 (2019).

[37] C. Zhang, Y. Zhang, X. Yuan, S. Lu, J. Zhang, A. Narayan, Y. Liu, H. Zhang, Z. Ni, R. Liu, E. S. Choi, A. Suslov, S. Sanvito, L. Pi, H.-Z. Lu, A. C. Potter, and F. Xiu, *Quantum Hall Effect Based on Weyl Orbits in Cd$_3$As$_2$*, Nature **565**, 331 (2019).

[38] C. M. Wang, H.-P. Sun, H.-Z. Lu, and X. C. Xie, *3D Quantum Hall Effect of Fermi Arcs in Topological Semimetals*, Phys. Rev. Lett. **119**, 136806 (2017).

[39] J. Yu and C.-X. Liu, *Piezoelectricity and Topological Quantum Phase Transitions in Two-Dimensional Spin-Orbit Coupled Crystals with Time-Reversal Symmetry*, Nature Communications **11**, 2290 (2020).

[40] J. I. A. Li, Q. Shi, Y. Zeng, K. Watanabe, T. Taniguchi, J. Hone, and C. R. Dean, *Pairing States of Composite Fermions in Double-Layer Graphene*, Nature Physics **15**, 898 (2019).

[41] M. Levin and M. P. A. Fisher, *Gapless Layered Three-Dimensional Fractional Quantum Hall States*, Phys. Rev. B **79**, 235315 (2009).

[42] H. Sakai, H. Fujimura, S. Sakuragi, M. Ochi, R. Kurihara, A. Miyake, M. Tokunaga, T. Kojima, D. Hashizume, T. Muro, K. Kuroda, T. Kondo, T. Kida, M. Hagiwara, K. Kuroki, M. Kondo, K. Tsuruda, H. Murakawa, and N. Hanasaki, *Bulk Quantum Hall Effect of Spin-Valley Coupled Dirac Fermions in the Polar Antiferromagnet BaMnSb$_2$*, Phys. Rev. B **101**, 081104 (2020).

**Figure captions**

**Fig. 1| Structure determination of BaMnSb₂. a**, The schematic showing the *I2mm* crystal structure of BaMnSb₂. **b**, Schematic of the Sb zig-zag chain layer from [001] and [010] axis formed due to orthorhombic distortion. **c**, Simulated (left panel) and experimental (right panel) HAADF-STEM images along the [100] and [010] zone axes. **d**, Magnified experimental HAADF-STEM image with superimposed atom structures highlighting the orthorhombic distortion. The red arrows points to the Sb layer and shows how Sb atoms in the zig-zag chain layer form dimmer-like structure in the STEM image taken along the [010]-zone axis. **e**, Optical SHG microscopy of polar domains. The polar axes are indicated by the light blue arrows. The imaging light polarization conditions for the ω and the 2ω frequencies are indicated by red and purple arrows. The crystallographic axes in each domain are indicated. The light polarization conditions for the inset image are orthogonal to the corresponding conditions for the main image, hence the reversal in contrast that is observed. **f**, SHG polarimetry experiments (circles) in domains A and B indicated in panel **e**, and the corresponding theoretical fits (solid lines) are shown for two different SHG polarization directions indicated in the inset schematic. Details of the fit based on *2mm* point group symmetry are given in the Method section.

**Fig. 2| Electronic band structure of BaMnSb₂. a**, The band structure from the first-principles calculation, with the inset showing the first BZ and red dots labelling the two gapped Dirac cones at $K_\pm$. **b**, the calculated Dirac band dispersion near $K_+$ and $K_-$, with the spin projection being color coded (red, spin-up; blue, spin-down). **c**, Schematic illustration of the orbital evolution at $K_+$. Here $p_\pm = (p_x \pm ip_y)/\sqrt{2}$; all orbitals come from Sb2 (see Fig. 1b) and the two bases in the black dashed box form the gapped Dirac cone. **d**, Constant energy contour of Ba(Mn₀.₉Zn₀.₁)Sb₂ on the $k_x$-$k_y$ plane, which is acquired by integrating the intensity from -15 meV to -5 meV. The ARPES data was taken with the photon energy of 30 eV. Two point-like hole pockets near $\overline{X}$ point can be seen. The presence of hole pockets near $\overline{Y}$ point is due to twin domains. **e**, ARPES spectrum along $\overline{\Gamma X}$ (cut 1 in panel **d**). There is only one crossing point near the Fermi level ($E_F$) at $\overline{X}/\overline{Y}$. **f**, ARPES spectrum along $\overline{MX}$ (cut 2 in panel **d**). Two crossing points at $E_F$ can be clearly resolved near $\overline{X}/\overline{Y}$. **g**, Comparison of the calculated band (dotted lines) and the band probed by ARPES along $\overline{MX}$ (cut 2 in panel **d**).

**Fig. 3| Bulk quantum Hall effect in BaMnSb₂. a**, Magnetic field dependences of in-plane resistivity ($\rho_{xx}$, blue) and Hall resistivity ($\rho_{xy}$, red) at $T$ = 1.4 K for the Hall bar sample B#1, measured with field perpendicular to the Sb-plane up to 40 T. Hall plateaus in $\rho_{xy}$ are clearly observed. **b**, Normalized inverse Hall resistivity ($\rho_{xy}^0/\rho_{xy}$) and in-plane resistivity ($\rho_{xx}$) versus $B_F/B$. $1/\rho_{xy}^0$ is defined as the step size between the first and the second plateaus of $1/\rho_{xy}$(see supplementary Fig. S3). $B_F$ is the frequency of the SdH oscillations in $\rho_{xx}$. **c**, Landau level fan diagram built from the SdH oscillations of $\rho_{xx}$ at 1.4 K. **d,** The second derivative, $-d^2\rho_{xx}/dB^2$, which is in phase with $\rho_{xx}$, clearly reveals the oscillation peak splitting near $B$ = 17 T. **e** and **f**, Field dependences of $\rho_{xx}$ and $\rho_{xy}$ at various temperatures ($T$ = 1.4, 3.8, 15, 30, 50 and 100 K), up to 90 T. The inset in **e** shows the SEM image of sample B#1 with Hall bar geometry fabricated by FIB.

**Fig. 4| Bulk quantum Hall effect in Eu-and Zn-doped BaMnSb₂. a**, Magnetic field dependences of $\rho_{xx}$ and $\rho_{xy}$ of the Hall bar sample E#1, measured with field perpendicular to the Sb-plane up to 64 T (see Fig. S7). The arrows indicate the oscillation peak splitting in $\rho_{xx}$. **B**, Magnetic field dependence of $\rho_{zz}$ at various temperatures ($T$ = 0.7, 4.1, 10, 20, 40, 80 and 150 K), measured on another Eu-doped sample (E#2) with a similar composition. The red arrows indicate the splitting in an oscillation valley. **c**, Temperature dependence of $\rho_{zz}$ at 58.5T for sample E#2. **d**, Magnetic field dependences of $\rho_{xx}$ and $\rho_{xy}$ at 2K for sample Z#1. The comparison between samples Z#1, B#1 and E#1-3 is shown in supplementary Tab. S1. **e**, Schematic illustration of the valley splitting due to the ***q***-quadratic and Zeeman terms. **f**, The LLs calculated from the effective model around $K_{\pm}$. LLs for $K_+$ and $K_-$ are presented as the blue and red lines, respectively. The black line is $E_F$. Bottom panel in **f**: The DOS at $E_F$ with Gaussian disorder broadening $\Gamma_0$ = 2meV (blue) and $\Gamma_0$ = 3meV (red) [33]. The numbers above the black arrows label the LL filling at DOS minima.

**Figure 1**

**Figure 2**

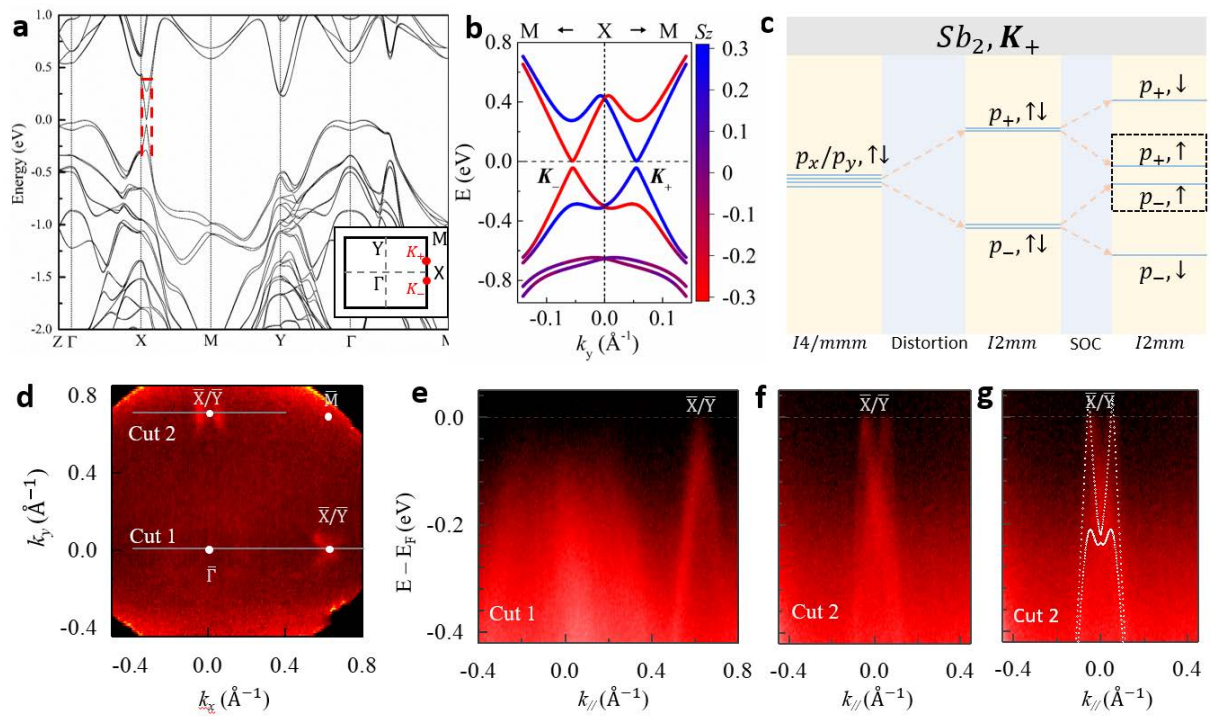

**Figure 3**

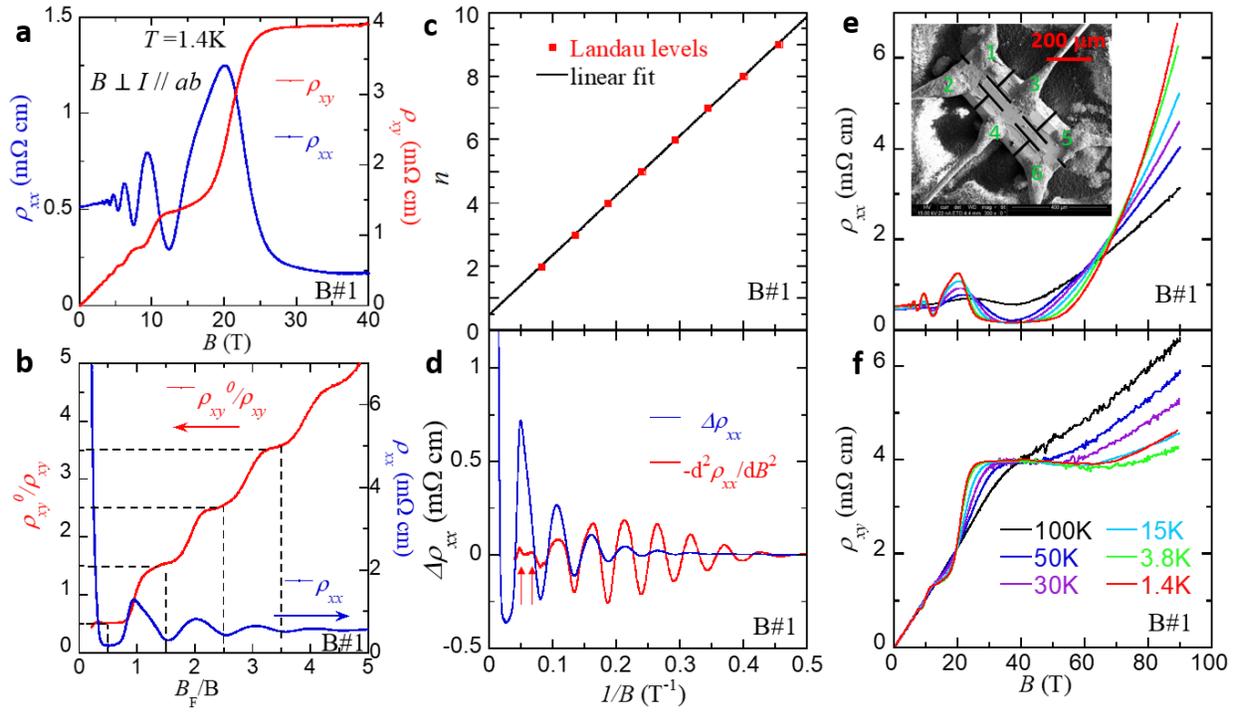

**Figure 4**

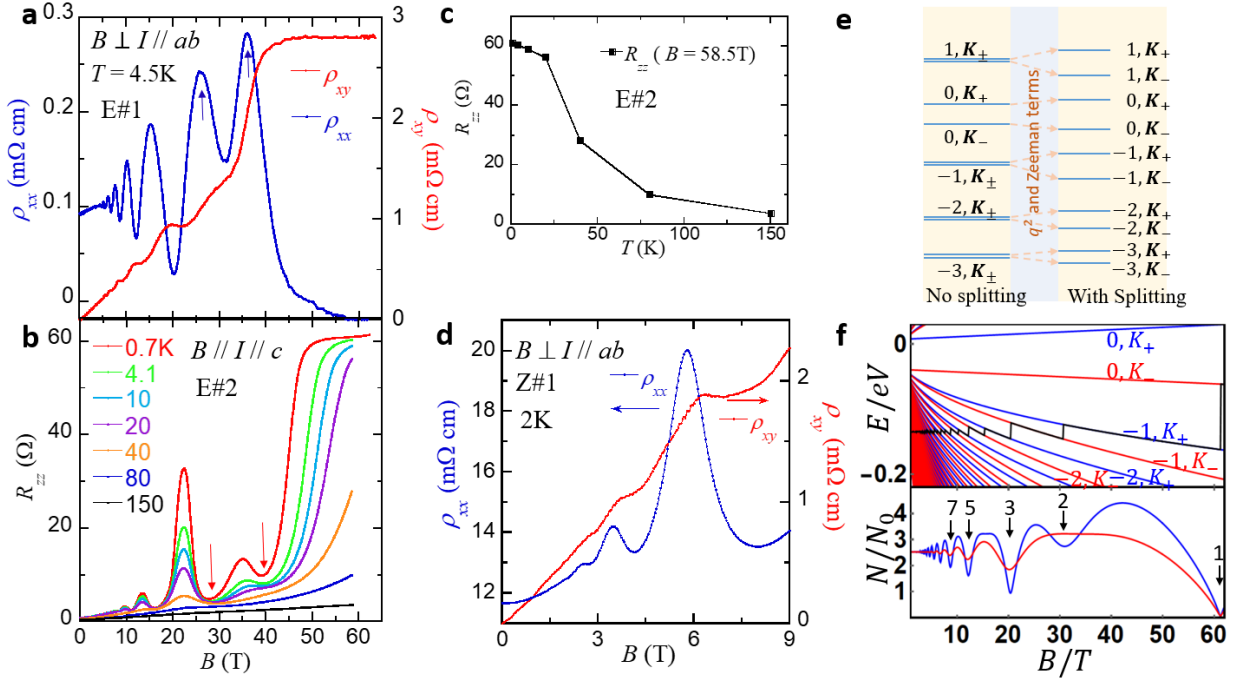